\documentclass[sigconf]{acmart}

\usepackage{graphicx}
\usepackage{multirow}
\usepackage{booktabs}
\usepackage[comments]{annotations}
\newcommand{\lc}[1]{\mynote{LC}{#1}{blue!20!magenta}}

\newcommand{\nop}[1]{}
\newcommand{\revision}[1]{{\color{red}{#1}}}
\newcommand{\Com}[1]{}

\usepackage[linesnumbered,algoruled]{algorithm2e}
\usepackage{algpseudocode}
\usepackage{amsthm}

\usepackage{cleveref}
\crefname{algocf}{alg.}{algs.}
\Crefname{algocf}{Algorithm}{Algorithms}
\SetKwRepeat{Do}{do}{while}

\usepackage{enumitem}

\usepackage{booktabs}
\usepackage{tabularx}
\newcolumntype{L}[1]{>{\raggedright\let\newline\\\arraybackslash\hspace{0pt}}m{#1}}
\newcolumntype{C}[1]{>{\centering\let\newline\\\arraybackslash\hspace{0pt}}m{#1}}
\newcolumntype{R}[1]{>{\raggedleft\let\newline\\\arraybackslash\hspace{0pt}}m{#1}}

\usepackage{caption}
\usepackage{subcaption}

\usepackage[all]{background}
\usepackage{stackengine}
\setstackEOL{\\}
\setstackgap{L}{\normalbaselineskip}
\SetBgContents{\color{gray}{\scriptsize \Longstack{PREPRINT - accepted at 59th ACM/IEEE Design Automation Conference (DAC) 2022}}}
\SetBgPosition{4.5cm,1cm}
\SetBgOpacity{1.0}
\SetBgAngle{0}
\SetBgScale{1.8}

\copyrightyear{2022} 
\acmYear{2022} 
\setcopyright{acmlicensed}\acmConference[DAC '22]{Proceedings of the 59th ACM/IEEE Design Automation Conference (DAC)}{July 10--14, 2022}{San Francisco, CA, USA}
\acmBooktitle{Proceedings of the 59th ACM/IEEE Design Automation Conference (DAC) (DAC '22), July 10--14, 2022, San Francisco, CA, USA}
\acmPrice{15.00}
\acmDOI{10.1145/3489517.3530543}
\acmISBN{978-1-4503-9142-9/22/07}

\begin{document}

\title[ALICE: An Automatic Design Flow for eFPGA Redaction]{ALICE: An Automatic Design Flow for eFPGA Redaction}
\author[C. Muscari Tomajoli, L. Collini, et al.]{
Chiara Muscari Tomajoli$^1$\textsuperscript{*}, Luca Collini$^1$\textsuperscript{*}, 
Jitendra Bhandari$^2$, Abdul Khader Thalakkattu Moosa$^2$, Benjamin Tan$^3$,
Xifan Tang$^4$, Pierre-Emmanuel Gaillardon$^4$,
Ramesh Karri$^2$, Christian Pilato$^1$}
\affiliation{\institution{$^1$Politecnico di Milano, Italy, $^2$New York University, USA, $^3$University of Calgary, Canada, $^4$University of Utah, USA}
\country{}
}

\begin{abstract}
Fabricating an integrated circuit is becoming unaffordable for many semiconductor design houses. Outsourcing the fabrication to a third-party foundry requires methods to protect the intellectual property of the hardware designs. Designers can rely on embedded reconfigurable devices to completely hide the real functionality of selected design portions unless the configuration string (bitstream) is provided. However, selecting such portions and creating the corresponding reconfigurable fabrics are still open problems. We propose ALICE, a design flow that addresses the EDA challenges of this problem. ALICE partitions the RTL modules between one or more reconfigurable fabrics and the rest of the circuit, automating the generation of the corresponding redacted design.
\end{abstract}

\maketitle

{
\renewcommand{\thefootnote}{\fnsymbol{footnote}}
\footnotetext[1]{C. Muscari Tomajoli and L. Collini contributed equally to this work.
}}

\section{Introduction}

Hardware Intellectual Property (IP) protection is becoming one of the most important concerns during Integrated Circuit (IC) design and manufacturing~\cite{9310331}. Due to the globalization of the supply chain, more semiconductor design houses are forced to outsource IC fabrication to third-party foundries to keep the costs sustainable. However, rogue employees can steal the IC design and make illegal copies~\cite{DBLP:journals/todaes/ShamsiLPFPJ19}. Design houses are using protections like \textit{watermarking}, \textit{split manufacturing}, and \textit{logic locking} to protect the critical parts of their designs~\cite{9310331}. All these methods have their limitations: watermarking is only a \textit{passive} method~\cite{survey_watermarking}; split manufacturing requires advanced manufacturing skills~\cite{9216128}, and logic locking is challenged by a broad range of attacks~\cite{DBLP:journals/corr/abs-2006-06806,DBLP:journals/todaes/ShamsiLPFPJ19}, especially when the attacker can access a working chip (called \textit{oracle}). 

\begin{figure}[t]
\centering
\includegraphics[width=0.78\columnwidth]{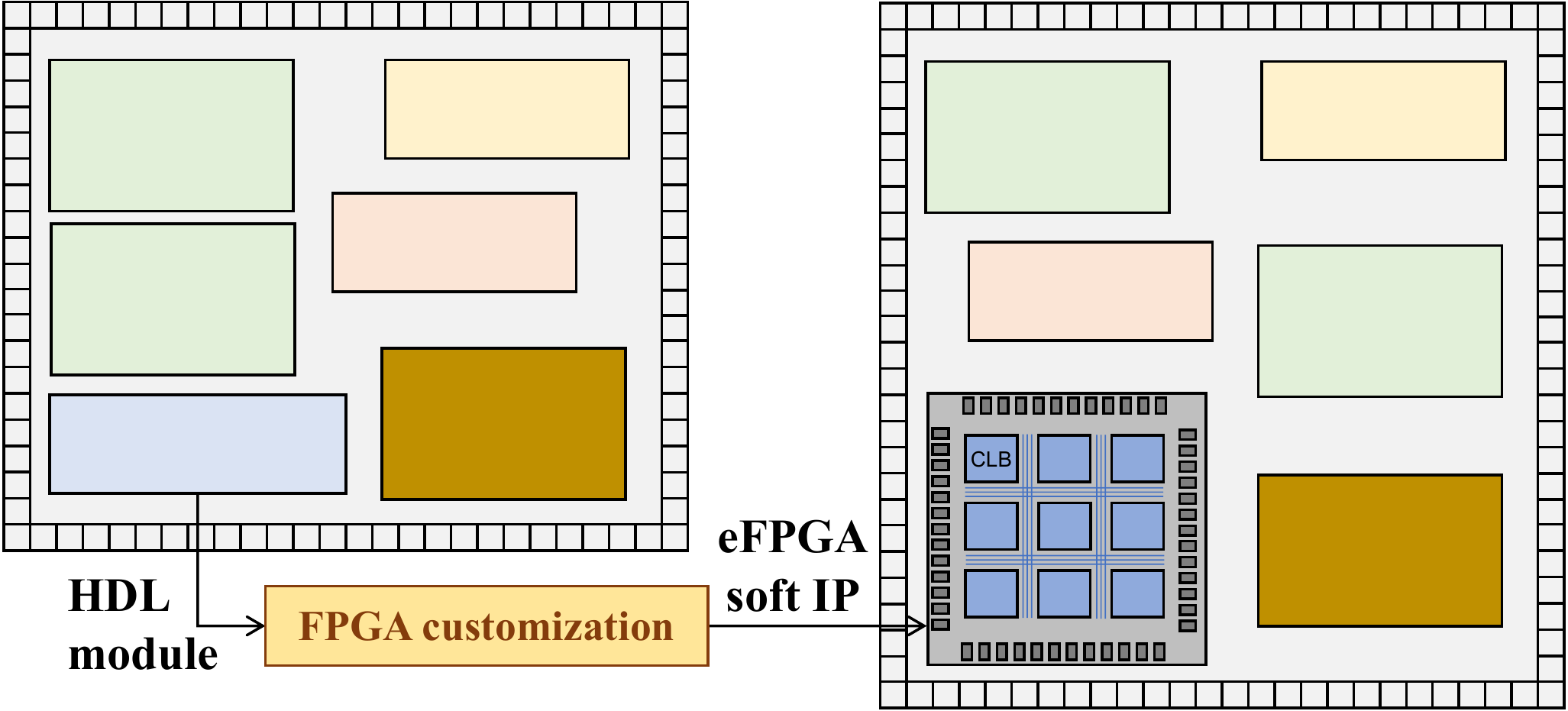}
\vspace{-8pt}\caption{FPGA redaction flow. Critical modules are replaced with custom eFPGA implementations.}\label{fig:efpga_flow}
\end{figure}

\textbf{FPGA redaction} is a novel, promising technique that aims to thwart reverse engineering attacks by exploiting the flexibility of reconfigurable devices. Critical parts are mapped on and replaced by specific reconfigurable blocks (called embedded FPGAs - eFPGAs) with a two-fold goal: (1) during fabrication, reconfigurable devices can implement any arbitrary functions, without revealing their intended functionality; (2) during execution, they can be configured to implement the correct functionality by classic FPGA programming methods. 
\Cref{fig:efpga_flow} shows an example, where a module is replaced by a custom eFPGA fabric. Inside, each block represents a Configurable Logic Block (CLB). Modern FPGA specialization tools, like OpenFPGA~\cite{9098028} and FABulous~\cite{DBLP:conf/fpga/KochDHYA21}, allow designers to start from a HDL module and generate the corresponding soft eFPGA IP that can be integrated and synthesized with the rest of the chip.
The FPGA-redaction resilience to SAT attacks comes from a large number of ``key bits'' to be recovered (i.e., the entire eFPGA configuration bitstream) and a more complex I/O relationship in the eFPGA fabric~\cite{mohan_hardware_2021,our_iccad_21}. Custom eFPGAs have smaller overhead than commercial, off-the-shelf ones~\cite{our_iccad_21,9369856}.

FPGA redaction requires the designers to perform several steps. First, it requires them to select the best modules to be redacted from both security and design viewpoints. Then, it requires the creation and integration of the corresponding custom eFPGA fabric. These two problems are strictly interdependent and often application-dependent. For these reasons, designers currently solve these by hand, potentially leading to sub-optimal solutions~\cite{9369856,Chen_dac_2020}.

\setlength{\textfloatsep}{8pt}

This paper focuses on the EDA problem of \textit{partitioning RTL modules} between eFPGA and ASIC and \textit{creating the proper eFPGA fabrics} to implement the redacted modules. While modules implemented in ASIC can be retrieved by the malicious foundry, the flexibility of eFPGAs protects the redacted modules.
We propose \textbf{ALICE} (\uline{A}utomatic module se\uline{L}ect\uline{I}on for se\uline{C}urity-aware \uline{E}FPGA redaction), a \textbf{complete flow to identify the modules to be redacted and generate the corresponding soft eFPGAs}. 
Starting from the set of candidate redaction modules, ALICE performs a progressive refinement of the solution by filtering out inadmissible modules, clustering the remaining ones to enable the creation of larger eFPGAs, and characterizing them in terms of hardware cost and security resilience to select the best final implementation.
After presenting the background of our work (threat model, eFPGA design flow, and related work), we present our main contributions:
\begin{itemize}[leftmargin=1.5em]
	\item we refine the list of modules to be redacted (\Cref{sec:selection});
	\item we group independent modules into clusters (\Cref{sec:identification}) and we characterize the corresponding eFPGA fabric (\Cref{sec:efpga_selection});
	\item we evaluate our automatic creation of FPGA-redacted designs on common benchmarks for hardware IP protection (\Cref{sec:results}).
\end{itemize}
Designers can combine functional characteristics (e.g., modules that affect selected outputs), structural characteristics (e.g., maximum number of I/O pins), and eFPGA parameters (e.g., maximum number of eFPGA instances) to guide the redaction process.  

\section{Background}

\subsection{Threat Model}

We assume the attackers have access to the chip design, can isolate the eFPGA fabric, and have access to an oracle, i.e., a fully-scanned and unlocked design.
In this way, they can observe input/output behaviors of this part to apply SAT-based attacks~\cite{7140252}. The attackers have to retrieve the correct bitstream to restore the real functionality. This is the typical threat model for recent eFPGA redaction works~\cite{our_iccad_21,mohan_hardware_2021}. In this scenario, \textbf{the eFPGA security comes more from the fabric parameters and way the designer uses the fabrics rather than the specific redacted modules themselves}~\cite{our_iccad_21,bhandari2021fabrics}. We also assume the designers will use state-of-the-art eFPGA parameters from the security viewpoint~\cite{bhandari2021fabrics}.

\subsection{Custom eFPGA Design Flow}

Reconfigurable devices can implement any arbitrary function after fabrication by simply changing the configuration bitstream. This is a key feature for hardware IP protection. Designers can integrate the FPGAs as pre-existing blocks in ASIC designs, while their configuration is done only by the final user. The function implemented on the FPGA is thus unknown to the foundry.

\begin{figure}
\centering
\includegraphics[width=0.9\columnwidth]{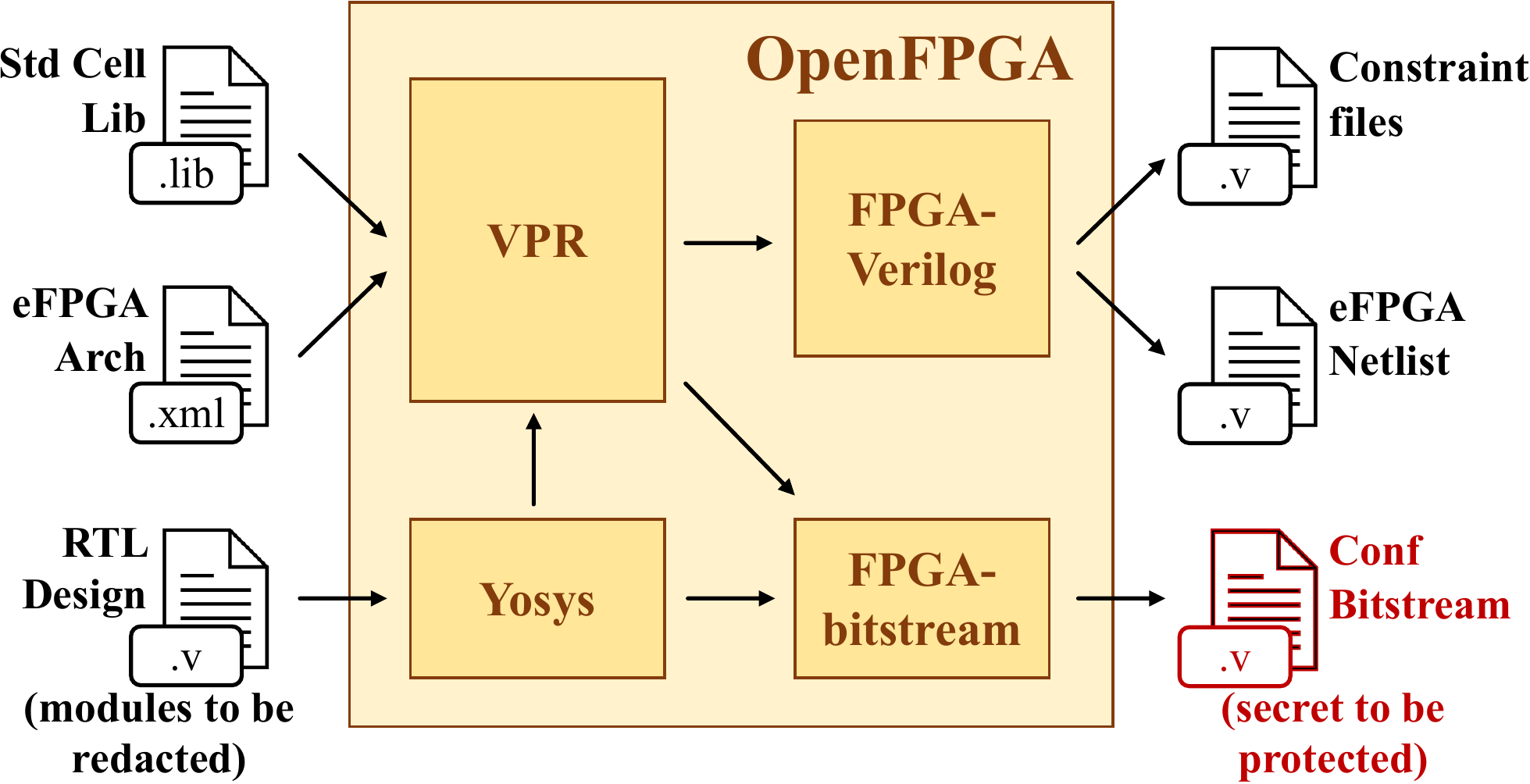}
\vspace{-6pt}\caption{ALICE uses an eFPGA design flow based on OpenFPGA~\cite{9098028}. The eFPGA netlist is integrated with the rest of the chip, while the configuration bitstream is kept secret.}\label{fig:fpga_flow}
\end{figure}

Custom eFPGAs can be created with open-source tools, like OpenFPGA~\cite{9098028} or FABulous~\cite{DBLP:conf/fpga/KochDHYA21}. Such frameworks allow the automatic customization of FPGA architectures, which are tailored to specific modules with a complete Verilog-to-bitstream flow. For example, \Cref{fig:fpga_flow} shows the OpenFPGA-based customization flow that can be used for eFPGA redaction~\cite{our_iccad_21}.  OpenFPGA starts from an XML specification of the fabric parameters and produces the corresponding fabrication-ready eFPGA IP~\cite{9098028,our_iccad_21}. The modules to be redacted will drive the customization of the eFPGA. Using open source frameworks offers additional degrees of freedom to the designer, where one can tune many parameters, as shown in~\cite{bhandari2021fabrics}. This will allow the user to come up with architectures that are most suitable for the given design. Thanks to a more tight integration of the soft eFPGA modules, the resulting System-on-Chip architectures can significantly reduce area and performance overheads~\cite{9369856,mohan_hardware_2021}.

In this work, we explore an FPGA architecture composed of Configurable Logic Blocks (CLBs) that are built with four 4-input LUTs, as proposed and evaluated in several recent works~\cite{9098028,DBLP:conf/fpga/KochDHYA21,our_iccad_21}. However, we can support any fabric configuration since our work is more focused on how to use them rather than in their generation and security evaluation. Indeed, we can support even off-the-shelf fabrics to be later integrated into the final chip.

\subsection{Related Work}

Hardware IP protection is a hot topic in recent years. Researchers proposed many methods, especially at low levels of abstraction (i.e., on gate-level netlists or physical designs, or directly during fabrication~\cite{8050883,eASIC}. For example, logic locking assumes the attacker is not able to retrieve the correct functionality thanks to the protection of a ``secret'', the locking key~\cite{8852678}. Despite many advances~\cite{8203496}, SAT attacks~\cite{7140252} can be used to identify the I/O relationships and retrieve key bits when an activated chip is available, challenging the effectiveness of logic locking~\cite{DBLP:journals/corr/abs-2006-06806,DBLP:journals/todaes/ShamsiLPFPJ19}.

FPGA redaction is a recent technique that aims at implementing selected modules with soft or hard eFPGAs that are included in the design. The key idea is that (1) attackers in the foundry have no access to the configuration of the bitstream that can implement any possible functionality, while (2) end-user attackers that have access to an activated chip cannot retrieve the correct bitstream. In this case, the ``secret'' corresponds to the configuration bitstream. However, the design of FPGA-redacted ICs is complex, especially in the module partitioning between eFPGA and ASIC. 

While recent studies focused on VLSI challenges of eFPGA integration~\cite{Mohan_fpga_2021}, the selection of the modules to be redacted is still manual effort or requires at least a reference design. In the former case, designers have to identify the modules to be protected, for example because they are part of the core business~\cite{mohan_hardware_2021}. Designers may want to use FPGA redaction to protect the results of selected outputs with FPGA redaction without knowing the critical components.
In the latter case, two or more designs are compared with each other to identify common parts (which are assumed to be common to many other designs) and different parts (which are the unique parts of the given design)~\cite{Chen_dac_2020}. However, designers may not have an alternative version of the same design to be compare with.

Recent studies on the security of FPGA redaction show that the resilience to SAT attacks is correlated more with the eFPGA fabric configuration and its utilization rather than the implemented module(s)~\cite{our_iccad_21,mohan_hardware_2021,bhandari2021fabrics}. For this reason, we focus more on the selection of the functionality to be redacted (together with its EDA implications), \textbf{assuming the fabric configuration as given and ``secure''}, and aiming at maximizing the fabric (both I/O and CLB) utilization.

ALICE performs the \textbf{automated FPGA redaction of a given design}, identifying the modules that have impact on selected outputs and enabling the possibility of grouping them into the same eFPGA to \textbf{maximize its utilization}. ALICE also supports \textbf{multiple eFPGA instances} to give more flexibility to the designer. 

\section{ALICE Design Flow for eFPGA Redaction}

\begin{figure}
\centering
\includegraphics[width=0.82\columnwidth]{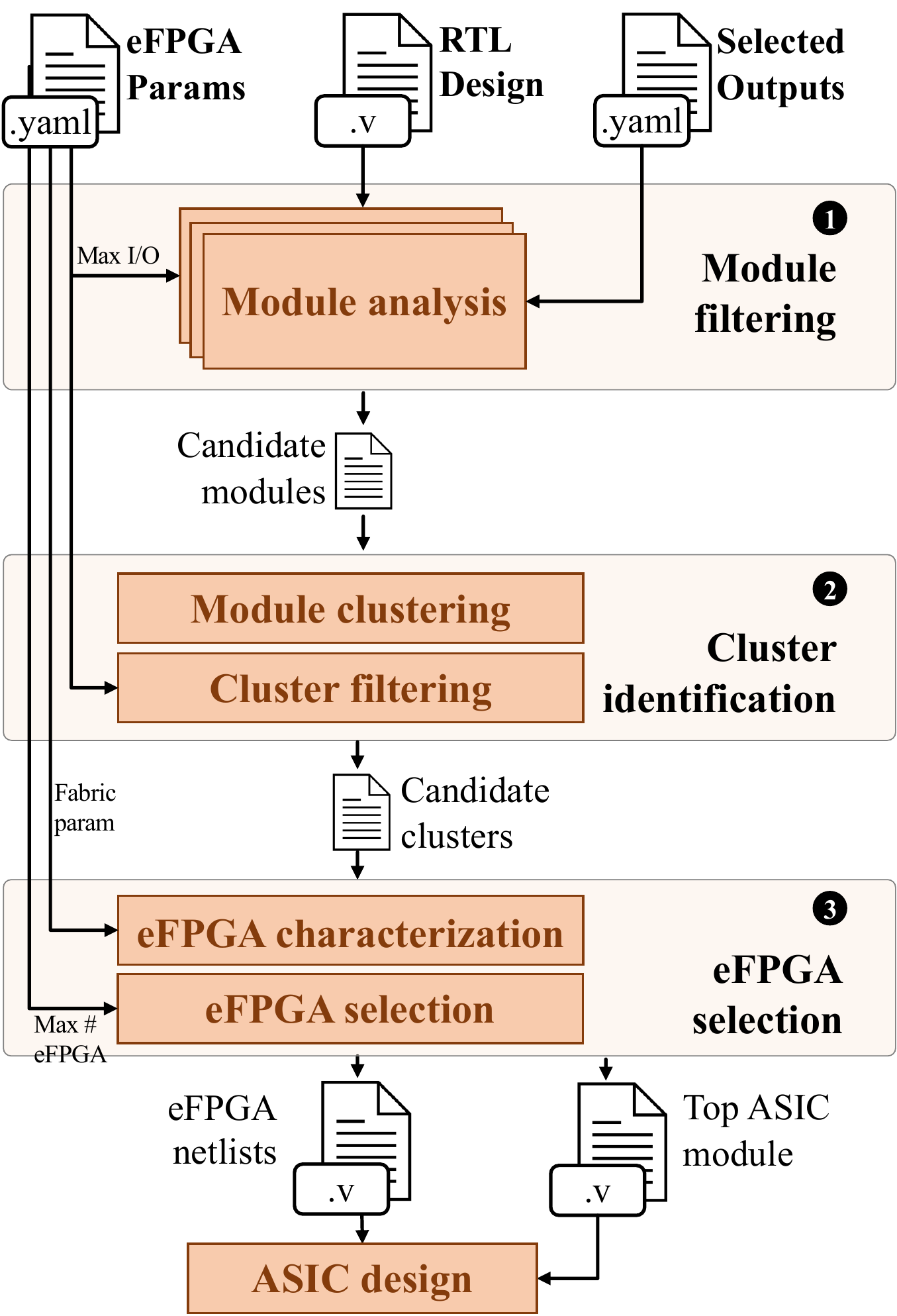}
\vspace{-4pt}\caption{ALICE flow for automatic eFPGA-based redaction.}\label{fig:flow}
\end{figure}

Our redaction flow is shown in Figure \ref{fig:flow}. It starts from the RTL description of the design to be redacted in Verilog\footnote{Limitations are only due to the HDL parser that we use. Supporting another HDL language (e.g., VHDL) only requires the proper parser.} and a set of parameters for the flow (in a custom YAML configuration file). Such parameters include eFPGA fabric configurations (e.g., as specified in the OpenFPGA configuration file), the maximum number of eFPGAs to be instantiated, and the maximum number of I/O for each of them. The number of I/O pins is also a rough indication of the type of eFPGA that the designer aims at using. For example, a 4$\times$4 fabric configuration has no more than 64 I/O pins~\cite{our_iccad_21,mohan_hardware_2021}. 
Currently we support (one or more) eFPGAs with the identical fabric architecture and maximum number of used I/O pins. While we contend that this setup will create a more regular physical design, adding support in the future for eFPGAs with different configurations is possible. 

ALICE focuses on how to partition an RTL design and generate the corresponding eFPGA-enhanced IC with three main phases: \textbf{module filtering}, \textbf{cluster identification}, and \textbf{eFPGA selection}. During module filtering, we analyze the design to identify \textit{candidate redaction modules}, while discarding the ones that do not satisfy specific constraints. In the second phase, the candidate modules are clustered into \textit{candidate module clusters}. Again, solutions that do not satisfy specific constraints are discarded. The result of this phase is a set of candidate clusters that are then characterized by running the flow for the creation of the corresponding eFPGA fabrics. We finally apply an algorithm to select the eFPGAs that maximize our objectives (i.e., minimum hardware overhead and maximum security) with no overlapping sets of redacted modules. The resulting redacted RTL description is reproduced along with the fabrics of the selected eFPGAs. The final output is the description of the final system ready for ASIC design.
ALICE is a modular flow that can be extended with additional criteria for selection. It can also interface with other eFPGA tools for characterization and include further metrics for security assessment, if needed.

\section{Module Filtering}\label{sec:selection}

This phase analyzes the input design to determine the list of RTL modules that must be considered for redaction. \Cref{alg:filtering} shows the pseudocode of our procedure. Our algorithm starts from the input design $D$, the list of eFPGA parameters $P$ (e.g., maximum number of I/O pins), and the list of selected output $O$. The designers can provide a list of outputs that they want to ``protect''. The algorithm then applies \textit{functional} and \textit{structural} criteria to obtain the final set $R$ of candidate redaction modules. Functional criteria aim at identifying modules that are more important for FPGA redaction from the functionality viewpoint. Structural criteria aim at identifying modules that can be effectively implemented with eFPGA, excluding the ones that would lead to an unfeasible solution.

We list the modules $M$ of the input design $D$ (line~\ref{l:extractmodules}), assigning an initial zero score for each of them (lines~\ref{l:initialize_score_b}-\ref{l:initialize_score_e}). We create the dataflow graph of the entire RTL design and, for each selected output, we increase the scores of the modules that have a direct impact on it (lines~\ref{l:score_b}-\ref{l:score_e}). We select the top-score modules and add them to the list $F$ of functionally-relevant modules for redaction (line~\ref{l:ranking}).

\begin{algorithm}[t] \footnotesize
	\DontPrintSemicolon 
	\KwIn{Input RTL design $D$, eFPGA parameters $P$, list of selected outputs $O$}
	\KwOut{Set of candidate redaction modules $R$}
    $M \gets \Call{ExtractInitialModules}{D}$\tcp*{Analyze input RTL design.}\label{l:extractmodules}
	$S \gets \emptyset$\;
    \ForEach {$m \in M $}{\label{l:initialize_score_b}
       $S[m] \gets 0$\;
    }\label{l:initialize_score_e}
    \ForEach {$o \in O $}{\label{l:score_b}
       $T \gets \Call{IdentifyModules}{M, o}$\tcp*{Compute modules $T$ affecting $o$}    

	   $\Call{UpdateScore}{T, S}$\tcp*{Increment scores of modules $T$}    
    }\label{l:score_e}
    $F \gets \Call{RankAndSelect}{M, S}$\tcp*{Select most relevant modules}\label{l:ranking}
    $R \gets \emptyset$\;
	\ForEach {$f \in F $}{\label{l:structural_b}
    	\uIf{$\Call{CheckParameters}{f, P}$}{\label{l:check_param}
    		$R \gets R \cup \{ f \}$\;\label{l:add_list}
    	}
    }\label{l:structural_e}
    \Return{$R$}	\;
	\caption{ALICE module filtering}
	\label{alg:filtering}
\end{algorithm}

In the next stage, we apply structural criteria to each functionally-relevant module for redaction (lines~\ref{l:structural_b}-\ref{l:structural_e}). We check whether each module is compatible with the given eFPGA parameters (line~\ref{l:check_param}). For example, we compute the number of I/O pins of the module to check if it fits into the potential eFPGA fabric. If the module satisfies the constraints, it can be added to the final list $R$ (line~\ref{l:add_list}).

The list $R$ represents feasible modules that affect a relevant number of (selected) outputs and can be clustered or implemented alone in an eFPGA (depending on their size). This phase can be easily extended with more module-level filtering criteria.

\section{Cluster Identification}\label{sec:identification}
\Com{
\textbf{Definition}: a module graph is a directed graph whose nodes are modules and whose edges represent instancing relationships between modules. If there is an edge from node A to node B, then module A creates and instance of module B. Consider a directed graph DG, DG is a module graph if each node represent a module and for each node N, for all modules M instantiated by module N, M is in DG. 

\textbf{Definition}: a cluster of modules is a disconnected graph whose subgraphs are all module graphs. 

\textbf{Clustering problem statement}: Given a set of candidate modules, a constraint on the eFPGA (we consider IO count so far) and a number of eFPGAs, find all the sets of distinct clusters that respect the constraint, the number of clusters in the set must be less or equal to the number of eFPGA and each cluster must respect the constraint on the eFPGA. Two clusters A and B are distinct \cp{?? please clarify} \revision{if there not exist an isomorphism between A and B} \lc{not sure if this is sufficient, but this would be concise}.
 
}
Given the set of candidate modules $R$, we find all valid combinations (\textit{clusters}) that can be redacted onto an eFPGA. A cluster can be composed of a single module (\textit{single-module redaction}) or a set of independent modules (\textit{multi-module redaction}). In both cases, the cluster is valid if the corresponding eFPGA implementation is admissible (i.e., it respects the given designer's constraints).

\begin{algorithm}[t] \footnotesize
	\DontPrintSemicolon 
	\KwIn{Set of candidate redaction modules $R$, eFPGA parameters $P$}
	\KwOut{Set of candidate module clusters $C$}
	$C \gets \emptyset $\;
	\ForEach {$r \in R $}{\label{l:initialize_cluster_b}
       $C \gets C \cup \{ r \}$\;
    }\label{l:initialize_cluster_e}
    $Flag \gets False$\;
    \Do{Flag}{\label{l:fpa_b}
			$D \gets \emptyset$\; 
	       	\ForEach {$c1 \in C $}{\label{l:recombine_b}
         	\ForEach {$c2 \in C $}{
         	\If{$c1 \neq c2$}{
         		$N \gets c1 \cup c2$\; 
	       		\If {$N \not\subset D  \wedge N \not\subset C \wedge \Call{CheckParameters}{N, P}$}{\label{l:check_cluster_b}
	    			$D \gets D \cup N$\;\label{l:add_cluster}
    			}
    		}
    		}
    	}\label{l:recombine_e}
		$Flag \gets False$\;
    	\If{$D \neq \emptyset$}{\label{l:check_new_cluster_b}
    		$C \gets C \cup D$\;
    		$Flag \gets True$\;
    	}
    }\label{l:fpa_e}
    \Return{$C$}\;
	\caption{ALICE cluster identification}
	\label{alg:identification}
\end{algorithm}

Algorithm~\ref{alg:identification} shows the pseudo-code of the procedure used in ALICE. It performs a \textit{fixed-point analysis} to identify the set $C$ of all candidate module clusters. Each of them is meant to fit into a single eFPGA. 
We initialize the set $C$ with clusters composed of the single modules identified in the previous phase (lines~\ref{l:initialize_cluster_b}-\ref{l:initialize_cluster_e}).
We then proceed iteratively to create new and larger clusters (lines~\ref{l:fpa_b}-\ref{l:fpa_e}). This part recombines each pair of admissible clusters (lines~\ref{l:recombine_b}-\ref{l:recombine_e}). If the cluster was not already identified in the previous iterations and it respects the input constraints (line~\ref{l:check_cluster_b}), it is added to the list of current clusters (line~\ref{l:add_cluster}).
Indeed, each cluster is analyzed with the same structural criteria used for the modules. For example, in the case of multi-module redaction, the number of I/O pins is the aggregated number of the I/O pins of the single modules. The cluster is admissible if it respects the limit given by the designer. 
At the end of each iteration, new clusters (line~\ref{l:check_new_cluster_b}) are added to the set $C$ and the procedure restarts. 
We terminate our algorithm when, after recombining the current clusters, it is not possible to create new ones.
Each element of $C$ is a candidate module cluster. 

\Com{Each element of the set $C$ is a list of candidate modules. We build the set $C$ iteratively. The procedure begins by inserting, in $C$ a list containing only $e \in M, \forall  e \in  M$ \$ . Then, for each element $c$ in $C$ we proceed by adding to $C$ all possible expansions of $c$ using the candidate modules in $M$. We add a new combination in $C$ only if it is not already present. A module $m$ can expand a combination $c$ if: (a) $\forall c' \in c, c'$ and $m$ are not related modules; (b) the combination obtained by adding m to c satisfies the constraints. If at least an expansion of $c$ has been found, we insert $c$ in an auxiliary set $D$ and remove it from $C$. We keep doing this expansion step until we do not expand any combination in an iteration. At the end of the procedure we do a union between $C$ and $D$  to get all the candidate module combinations.}

\section{eFPGA Selection}\label{sec:efpga_selection}

Each candidate module cluster in $C$ can be implemented by an eFPGA. The set of resulting candidate implementations must be now characterized, ranked, and selected to determine the final solution. In this phase, we evaluate all candidate clusters to determine whether the corresponding eFPGA fabrics are admissible, determine all feasible solutions, and select the best and final one. 

\begin{algorithm}[t] \footnotesize
	\DontPrintSemicolon 
	\SetKwProg{Fn}{Function}{:}{}
  	\KwIn{Set of candidate module clusters $C$, eFPGA parameters $P$}
	\KwOut{Solution $s_t$}
	$F \gets \emptyset$\;
	\ForEach {$c \in C $}{\label{l:char_openfpga_b}
       $f \gets \Call{CreateEFPGA}{c, P}$\;\label{l:runopenfpga}
       \If{$\Call{IsValid}{f}$}{\label{l:checkvalid}
       	   $F \gets F \cup f $\;\label{l:keepefpga}
	   }
    }\label{l:char_openfpga_e}
    $T \gets \Call{ComputeScore}{F}$\;\label{l:scoreefpga}
    $W \gets \{ \}$\tcp*{Initialize with empty solution}\label{l:initialize_empty}
    $S \gets \emptyset$\;
    \ForEach{$w \in W$}{\label{l:bb_b}
    	\ForEach{$f \in F$}{\label{l:addnewefpga_b}
    		$c \gets f \cup w$\;\label{l:createsolution}
    		\If{$\Call{isValidSolution}{c}$}{\label{l:checkvalidsol}
    			\If{$\Call{isFinal}{c}$}{\label{l:ifleaf}
    				$S \gets S \cup c$\;\label{l:addsolution}
    			}
    			\Else{
    				$W \gets W \cup c$\;\label{l:addworking}
    			}
    		}
    	}\label{l:addnewefpga_e}
    }\label{l:bb_e}
    $S \gets S \cup W \setminus \{ \}$\;
    $s_t \gets \Call{RankAndSelect}{S, T}$\;\label{l:rankselect}
    \Return{$s_t$}\;
	\caption{ALICE eFPGA selection}
	\label{alg:selection}
\end{algorithm}

Algorithm~\ref{alg:selection} shows the pseudocode used in ALICE. First, we  generate the top module corresponding to each candidate cluster and run the selected eFPGA customization flow (i.e., OpenFPGA in our case) on it (lines~\ref{l:char_openfpga_b}-\ref{l:char_openfpga_e}). In the case of multi-module redaction, we create a top Verilog module that instantiates all independent modules. OpenFPGA returns the corresponding fabric if the design is feasible and an error otherwise (e.g., when the cluster modules cannot be implemented for any reason). Since the designer can specify the range of permitted fabric sizes, we also check that the resulting fabric is admissible (line~\ref{l:checkvalid}). If so, the fabric is added to the list $F$ of valid implementations (line~\ref{l:keepefpga}). At this point, we give a score to each fabric implementation (line~\ref{l:scoreefpga}). The score combines information about I/O and CLB utilization as follows:
\begin{equation}\label{eq:score}
	T_f = \alpha\cdot\frac{MaxIOUtil - IOUtil_f}{MaxIOUtil}+\beta\cdot\frac{MaxCLBUtil - CLBUtil_f}{MaxCLBUtil}
\end{equation}
where $IOUtil_f$ and $CLBUtil_f$ represent the I/O and CLB utilization, respectively, while $MaxIOUtil$ and $MaxCLBUtil$ represent the corresponding maximum values for all analyzed eFPGAs. In this way, both contributions range between 0 and 1. $\alpha$ and $\beta$ are two user-defined parameters to balance the contributions.
The score $T$ embeds information related to security resilience. Indeed, eFPGA implementations with poor I/O utilization are more prone to attacks because it is easier to identify stuck-at-0 outputs. Similarly, fabrics with low CLB utilization have less logic to be (successfully) recovered~\cite{our_iccad_21,bhandari2021fabrics}.
We then use a \textit{branch\&bound algorithm} to enumerate all possible eFPGA combinations that can be redacted together (lines~\ref{l:bb_b}-\ref{l:bb_e}) and obtain the full set of solutions. In particular, we start from an empty working solution (line~\ref{l:initialize_empty}) and, at each step, we aim to add a new eFPGA implementation to each current working solution (lines~\ref{l:addnewefpga_b}-\ref{l:addnewefpga_e}). A solution represents a set of eFPGAs with no overlapping module instances. If the solution is final (i.e., it reaches the maximum number of allowed eFPGAs or it redacts all the admissible modules), it is added to the final set of solutions (line~\ref{l:addsolution}). Otherwise, we keep it in the working list for further expansion (line~\ref{l:addworking}). At the end of this phase, the set $S$ contains the total set of admissible solutions. 
We now assign a score to each of them. The score of a solution is the sum of the scores of its eFPGA implementations, each of them obtained with Eq.~\ref{eq:score}. We rank the set $S$ according to the score and the one with the highest score is the best and final solution (line~\ref{l:rankselect}). 

The final solution includes a set of eFPGA implementations, each of them containing a list of module instances to be redacted. At this point, we need to regenerate the top module for ASIC implementation (\textit{Top ASIC module} in Figure~\ref{fig:flow}) where we replace the redacted instances with the corresponding eFPGA instances. In case of multi-module redaction, the different modules can be spread around the design. In this case, we apply a \textit{dominator tree} analysis on the module hierarchy to identify the best point where to insert the eFPGA instance and minimize wire length. Signals from the original instances are re-routed to the corresponding eFPGA instance, while its control signals are propagated to the top module. We also remap the module signals to the eFPGA GPIO signals for correct connection. The updated design, along with the fabric netlists, can be given to physical design tools.

\section{Experimental Evaluation}\label{sec:results}

We implemented a prototype of ALICE in Python, using the PyVerilog framework~\cite{Takamaeda:2015:ARC:Pyverilog}. PyVerilog can parse the Verilog designs, analyze, and manipulate the resulting Abstract Syntax Tree (AST), and regenerate the output files, including the ones fed into the OpenFPGA tool chain for eFPGA creation.
Table~\ref{tab:bench} shows the benchmarks that we used to validate ALICE. They are commonly used to evaluate RTL locking~\cite{9427060}. The table reports the number of modules and the instances that can be redacted. We report the range of the I/O pin count for such modules. For each design, we identified the main output(s) to be given to the module filtering phase.

We configure ALICE to run with two configurations. In \texttt{cfg1}, we set the maximum I/O pin count of the modules that can redacted to 64 and the limit is two eFPGAs. In \texttt{cfg2}, the maximum I/O pin count is 96 and the limit is one eFPGA. These experiments will show how to use ALICE to implement more but smaller eFPGAs or fewer but larger eFPGAs. We set $\alpha=\beta=1$ (Eq.~\ref{eq:score}) in both cases.

We run the OpenFPGA flow to implement the eFPGA fabrics composed of 4-input fracturable LUTs, 4 logic elements for each CLB, and 8 GPIOs for each I/O tile. Future work will explore these eFPGA parameters. Each OpenFPGA run aims at identifying the most suitable fabric (i.e., the one with minimum size) to implement the given module(s).
We finally validated the designs with Cadence Genus 18.14 for logic synthesis and Cadence Innovus 18.10 for physical design, targeting the NanGate 45nm Open Cell Library.

\begin{table}[t]
\caption{Characteristics of the selected benchmarks}\label{tab:bench}
\centering
\vspace{-8pt}\begin{tabular}{@{}llcccl@{}} 
\toprule
Suite                   & Design   & Modules  & Instances & I/O pins  \\ 
                   &    & (\#)   & (\#) & [min, max]  \\ 
\midrule
\multirow{4}{*}{CEP~}   & DES3     & 11 & 11 & [12, 301]       \\
                        & FIR      & 5  & 5 & [64, 384]         \\
                        & IIR      & 5  & 5 & [66, 384]         \\
                        & SHA256   & 3  & 3 & [38, 774]        \\ 
\hline
\multirow{2}{*}{IWLS05} & SASC     & 2  &  3 & [23, 28]        \\
                        & USB\_PHY & 3  & 3 & [17, 33]      \\ 
\hline
\multirow{1}{*}{OpenROAD} & GCD    & 10  & 11 & [6, 68]         \\
\bottomrule
\end{tabular}
\end{table}

\begin{table*}[t]
\caption{Results after running ALICE with two different configurations.}\label{tab:results}
\centering
\vspace{-12pt}\begin{tabular}{@{}L{1.9cm}lcccccccccc@{}} 
\toprule
Configuration           & Design   & \# Instances & \multicolumn{2}{c}{Module filtering}  & \multicolumn{2}{c}{Cluster identif.} & \multicolumn{5}{c}{eFPGA selection}  \\ 
\cmidrule(l){4-5} \cmidrule(l){6-7} \cmidrule(l){8-12}
                        &          & & Time & $|R|$ & Time & $|C|$ & Time & \# valid & $|S|$ & eFPGA size & \# redacted \\
                        & & & & & & & & eFPGAs & & & modules \\ 
\hline
\multirow{8}{*}{\shortstack[l]{\texttt{cfg1}: \\64 I/O pins\\ and 2 eFPGAs}}   & DES3     & 11   & 289.21s & 8 & 0.96s & 218 & 905.12s & 216 & 2,105 &  8$\times$8, 8$\times$8   & 4   \\
                        & FIR      & 5 & 0.17s  & 1 & <0.01s & 1 & 1.43s & 1 & 1 & 6$\times$6 & 1  \\
                        & IIR      & 5   & 1.36s & 0  & - & - & - & - & (n.a.)\textsuperscript{1} & - & -  \\
                        & SHA256   & 3   & 12.87s & 1 &  <0.01s  & 1 &  4.80s  & 1 & 1 & 12$\times$12 & 1\\ 
						& SASC     & 3   & 0.20s &  1 & <0.01s & 1 &  1.36s & 1 & 1 &  7$\times$7 & 1     \\
                        & USB\_PHY & 3   &  2.03s     & 2 & <0.01s & 3 & 5.99s  & 1 & 1 & 7$\times$7 & 1\\ 
                        & GCD      & 11  &  0.39s  & 9 &  0.01s      & 28 & 32.10s & 19 & 76 & 4$\times$4, 4$\times$4 & 2\\ 
\midrule
\multirow{8}{*}{\shortstack[l]{\texttt{cfg2}: \\96 I/O pins\\ and 1 eFPGA}}   & DES3     & 11   & 295.65s & 8 & 1.17s & 255 & 1,093.03s & 255 & 245 & 14$\times$14 & 8       \\
                        & FIR      & 5   & 0.15s & 3 & <0.01s & 3 & 16.76s & 3 & 3 & 6$\times$6   & 1\\
                        & IIR      & 5   & 0.29s  & 2 & <0.01s  & 2 &  24.40s  & 2 & 2 & 15$\times$15 & 1\\
                        & SHA256   & 3   & 12.68s  & 1 &  <0.01s  & 1 & 4.83s & 1 & 1 & 12$\times$12 & 1\\ 
						& SASC     & 3   & 0.20s & 1 & <0.01s & 1 & 1.34s & 1 & 1 &  7$\times$7 & 1 \\
                        & USB\_PHY & 3   & 1.92s & 2 & <0.01s & 3 & 5.77s & 1 & 1 & 7$\times$7 & 1\\ 
                        & GCD      & 11  & 0.45s & 10 &  0.05s & 70 & 91.28s & 37 & 33 & 5$\times$5 & 3\\ 
\bottomrule
\end{tabular}\\
\footnotesize{\textsuperscript{1}The module with the minimum I/O count already exceeds the maximum I/O size of the eFPGA (see Table~\ref{tab:bench}).}\vspace{-6pt}
\end{table*}

\begin{figure}[t]
 \begin{subfigure}[b]{0.47\columnwidth}
 \centering
 \includegraphics[height=3.35cm]{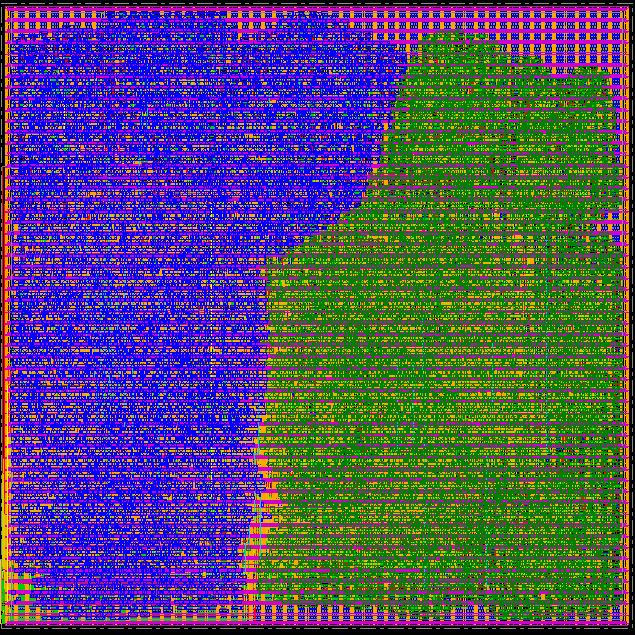}
 \caption{\texttt{cfg1}: two 4$\times$4 (52,629$\mu m^2$)}\label{fig:gcd_2}
 \end{subfigure}
 \hfill
 \begin{subfigure}[b]{0.47\columnwidth}
 \centering
 \includegraphics[height=3.47cm]{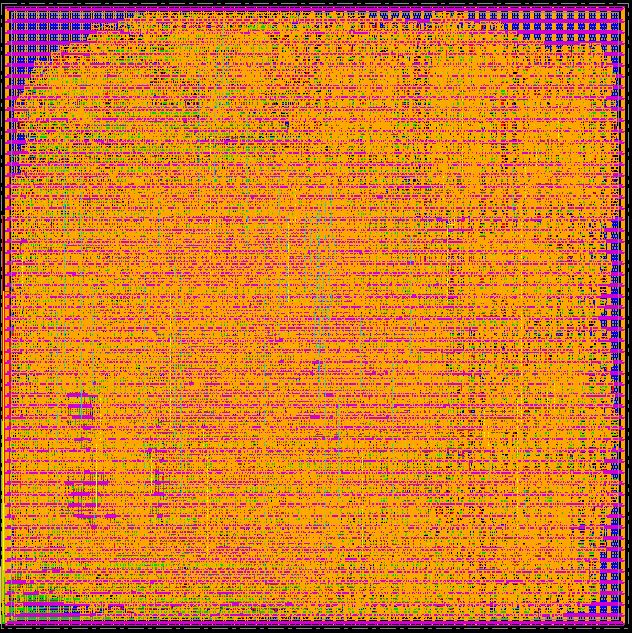} 	
 \caption{\texttt{cfg2}: one 5$\times$5 (54,512$\mu m^2$)}\label{fig:gcd_1}
 \end{subfigure}
\vspace{-3pt}\caption{Physical layouts of two GCD solutions with different number of eFPGAs. The figures are in scale.}\label{fig:layouts}	
\end{figure}

Table~\ref{tab:results} shows the results that we obtained when running ALICE on the benchmarks with the two configurations. In particular, we report: the number of candidate redaction modules ($|R|$), as obtained after \textit{module filtering}, the number of candidate module clusters that are created ($|C|$), as obtained after \textit{cluster identification}, the total number of valid eFPGA implementations, and the number of total solutions ($|S|$), as obtained during \textit{eFPGA selection}. We also report the characteristics of the eFPGAs in the final solution, along with the total number of redacted modules. 

In all benchmarks except IIR in the first configuration, we are able to find at least a feasible solution for the given eFPGA parameters. In the case of IIR in \texttt{cfg1}, the smallest modules have already more I/O pins (66 pins -- see Table~\ref{tab:bench}) than the maximum allowed I/O pin count of the eFPGA (64 pins). Indeed, the filtering phase does not produce any valid candidate redaction module and the flow cannot continue. Increasing the number of I/O pins to 96 (\texttt{cfg2}) allowed us to find a solution, showing how \textbf{ALICE can guide the designer in the identification of modules to be redacted}.

DES3 and GCD present other interesting results. They have more instances than the other benchmarks and ALICE is able to find several candidate clusters in both configurations (more than 200 for DES3 and at least 19 for GCD). Both GCD and DES3 have modules with a high variance in the number of I/O pins. So we can create multi-module redaction clusters when combining modules with low and high number of I/O pins, but clusters having modules with many I/O pins become invalid. Also, when using more eFPGAs (\texttt{cfg1}), the number of possible cluster combinations and, in turn, solutions grows significantly. The tool is then able to find solutions with two smaller eFPGAs (two 8$\times$8 for DES3 and two 4$\times$4 for GCD) or one larger eFPGA (14$\times$14 for DES3 and 5$\times$5 for GCD). The designers can use these results in different ways. For DES3, they can use the second implementation with a 14$\times$14 eFPGA because it redacts many more modules than the first case. For GCD, the two solutions are equivalent from the area viewpoint (see data in Figure~\ref{fig:layouts}) and have almost the same number of redacted modules. The designer could be motivated to use the first solution because it requires the attacker to recover more bit-streams. Figure~\ref{fig:layouts} shows screenshots of the two physical designs for GCD. This testcase is small and so most of the chip is occupied by the eFPGAs. However, the same modules will become less relevant when the component is inserted into a larger system-on-chip (like PicoSoc in~\cite{our_iccad_21}). In all cases, \textbf{area/time/power overheads are in line with previous studies on FPGA redaction~\cite{our_iccad_21,mohan_hardware_2021} as they are more related to the fabric architectures and sizes rather than the specific modules that are redacted}.

{\balance
Table~\ref{tab:results} reports also the execution time of each phase. Note that \textit{module filtering} includes the time spent for dataflow analysis to determine which modules affect the selected outputs, while \textit{eFPGA selection} includes the time to run OpenFPGA on all valid solutions.
Dataflow becomes more complex as the complexity (and not necessarily the number) of the RTL modules increases and is mostly independent of the configuration parameters. For example, SHA256 takes more time than GCD even if it has only three modules.
Running OpenFPGA is generally very fast, except for large eFPGA instances. However, it is always in the order of tens of seconds in the worst case (see IIR with two large solutions in \texttt{cfg2}). In general, the time for the \textit{eFPGA selection} phase grows linearly with the number of solutions to be tested.
In FIR, even if the solutions for the two configurations are the same, the \texttt{cfg2} analysis takes much longer because there are more and larger eFPGA candidates (up to 19$\times$19) that are later excluded because of low score.

\section{Conclusions and Future Work}

This paper presented ALICE, a methodology for automatic eFPGA redaction. ALICE analyzes the given RTL design, identifies the modules that have an effect on selected outputs, and cluster them. Such clusters are then characterized with an open-source FPGA customization flow and we select the ones that can maximize security. We show that ALICE can identify different solutions based on the given parameters (e.g., maximum number of I/O pins and eFPGA instances). 
Our flow can be part of a larger exploration that co-optimizes eFPGA parameters and module selection. It can also include a pre-processing step to perform fine-grained redaction: It can decompose large modules into smaller instances so that only part of them are effectively redacted.

\section*{Acknowledgements}
R. Karri was supported in part by ONR Award \# 
N00014-18-1-2058, NSF Grant \# 1526405, NYU Center for Cybersecurity, and NYUAD
Center for Cybersecurity. X. Tang and P.-E. Gaillardon were supported in part by DARPA, under the grants \# FA8650-18-2-7855 and \# FA8650-18-2-7849.\vspace{-0.5em}

\bibliographystyle{ACM-Reference-Format}
\bibliography{main.bib}
}

\end{document}